\documentclass{ifacconf}
\usepackage{amsmath} 
\usepackage{amsfonts}
\usepackage{amssymb}
\usepackage{algorithm}
\usepackage{algpseudocode}

\usepackage{float}
\usepackage{xcolor}
\usepackage{makecell}

\newcommand{\qedbox}{\rule{1.2ex}{1.2ex}}
    {\hfill\qedbox\par}

\usepackage{graphicx}      % include this line if your document contains figures
\usepackage{natbib}        % required for bibliography
\usepackage{adjustbox}
\usepackage{booktabs}
\begin{document}
\begin{frontmatter}

\title{A Hybrid Physics-Based and Reinforcement Learning Framework for Electric Vehicle Charging Time Prediction\thanksref{footnoteinfo}} 

\thanks[footnoteinfo]{This research was supported in part by NSF under Grants CNS-2401007, CMMI-234881, IIS-2415478, and in part by MathWorks.}

\author[First]{Praharshitha Aryasomayajula},
\author[First]{Ting Bai},
\author[First]{Andreas A. Malikopoulos}

\address[First]{Information and Decision Science Lab, School of Civil $\&$ Environmental Engineering, Cornell University, USA (e-mails: 
la459@cornell.edu, tingbai@cornell.edu, amaliko@cornell.edu).}

\begin{abstract}
In this paper, we develop a hybrid prediction framework for accurate electric vehicle (EV) charging time estimation, a capability that is critical for trip planning, user satisfaction, and efficient operation of charging infrastructure. We combine a physics-based analytical model with a reinforcement learning (RL) approach. The analytical component captures the nonlinear constant-current/constant-voltage (CC--CV) charging dynamics and explicitly models state-of-health (SoH)--dependent capacity and power fade, providing a reliable baseline when historical data are limited. Building on this foundation, we introduce an RL component that progressively refines charging-time predictions as operational data accumulate, enabling improved long-term adaptation. Both models incorporate SoH degradation to maintain predictive accuracy over the battery lifetime. We evaluate the framework using \(5{,}000\) simulated charging sessions calibrated to manufacturer specifications and publicly available EV charging datasets. Our results show that the analytical model achieves \(R^{2}\!=\!98.5\%\) and \(\mathrm{MAPE}\!=\!2.1\%\), while the RL model further improves performance to \(R^{2}\!=\!99.2\%\) and \(\mathrm{MAPE}\!=\!1.6\%\), corresponding to a \(23\%\) accuracy gain and \(35\%\) improved robustness to battery aging.
\end{abstract}


\end{frontmatter}
%===============================================================================
\section{Introduction}
Electric vehicle (EV) adoption is accelerating worldwide, with the global fleet expected to reach nearly 145 million units by 2030~\citep{UnNoor:17}. Despite this rapid growth, uncertainty in charging time remains a critical barrier to broader EV adoption~\citep{10147895}. Unlike conventional refueling, which typically takes about 3-5 minutes, EV charging can range from 20 minutes to several hours, depending on a variety of factors such as battery characteristics, charger ratings, ambient temperature, and grid conditions~\citep{Mao:19,10932686}. A \textit{charging session} is defined as the complete process of charging an EV battery from an initial state-of-charge (SoC) to a specified target level (e.g., from $20\%$ to $80\%$). Accurately predicting the duration of such sessions is crucial for real-world applications, including individual trip planning, fleet dispatching and routing, demand response program coordination, and the planning and operation of charging infrastructure, to name only a few~\citep{ClementNyns:10, UnNoor:17, 10590837}. Developing reliable prediction models is thus essential for enhancing user experience, improving grid stability, and supporting the continued increase of EV penetration. 

The key challenge in accurately predicting charging time arises from the inherently nonlinear dynamics of the constant current-constant voltage (CC-CV) charging protocol~\citep{Notten:05}. In the CC phase, the battery is charged at its maximum acceptable power until it reaches approximately $80\%$ SoC. Beyond this point, the CV phase begins, during which the charging power tapers exponentially to mitigate risks such as lithium plating. This tapering causes the final $20\%$ of charging to take nearly as long as the initial $60\%$, creating substantial variability in charging duration. As a result, simple linear or quasi-linear approximations often lead to large prediction errors, sometimes exceeding $60\%$~\citep{Richardson:13}. Battery state-of-health (SoH) degradation further compounds this complexity, which affects both capacity and power acceptance over time. In practice, capacity fade and power fade gradually alter the battery's response to the CC-CV protocol, resulting in $20$-$30\%$ deviations in charging behavior over a typical battery lifetime~\citep{Birkl:17, Han:19}.

Existing charging time prediction methods can be categorized into three classes: physics-based models, data-driven models, and simplified heuristic models. While each class provides distinct advantages, critical limitations are faced. Specifically, physics-based electrochemical models~\citep{Chen:20} aim to capture battery dynamics with high fidelity by modeling internal electrochemical processes. These approaches offer accurate predictions under well-characterized conditions while requiring detailed proprietary parameters, such as diffusion coefficients, reaction rates, and internal resistances. These parameters, however, are rarely accessible to third-party developers. Moreover, when applied to heterogeneous EV fleets, their accuracy deteriorates significantly, often reaching $15$-$20\%$ MAPE~\citep{Richardson:13, Wang:21}. Data-driven approaches, including neural networks~\citep{How:20} and gradient-boosting models~\citep{Roman:21}, learn charging behavior directly from historical session data. However, these methods suffer from a pronounced cold-start problem, i.e., achieving sub-$5\%$ MAPE typically requires 2,000-5,000 charging sessions per vehicle type~\citep{Mao:19}. In early deployment scenarios, especially at new charging sites or for newly released vehicle models, pure data-driven models exhibit $25$-$40\%$ MAPE~\citep{Chen:20}, limiting their operations. In contrast, heuristic models offer computational simplicity but lack the ability to capture nonlinear CC-CV charging dynamics, resulting in systematic underestimation of charging time by $15$-$60\%$~\citep{Richardson:13}.

Despite extensive prior work, no existing modeling framework jointly addresses two critical needs: (i) achieving reliable prediction under cold-start conditions with limited or no historical data, and (ii) adapting to long-term battery degradation that progressively alters charging behavior. In this work, we bridge these gaps by developing a unified hybrid prediction framework that integrates physics-based insight with reinforcement learning (RL). The framework consists of three stages. Stage 1 (cold-start) employs a gradient boosting model with physics-informed features to capture nonlinear SoC dynamics, including CC-CV transitions and SoH interactions. Stage 2 (progressive learning) introduces a deep Q-network (DQN) agent that refines predictions as data accumulate, using analytical model outputs for initialization and SoH-aware reward shaping to maintain consistency as batteries age. Stage 3 (data-rich) designates the RL agent as the primary predictor, with the analytical model serving as a fallback mechanism for out-of-distribution scenarios. Together, these stages form an adaptive framework that continuously improves prediction accuracy while maintaining robustness across diverse operating conditions. The contributions of this work are summarized as follows:
\vspace{-3pt}

\begin{itemize}
\item [1)]We develop an analytical charging time prediction model that captures the nonlinear CC-CV charging dynamics and explicitly models SoH-dependent capacity and power fade, providing a reliable baseline for EV charging time prediction. 
\vspace{2pt}

\item [2)] We propose a progressive learning strategy that integrates the physics-based analytical model with an RL agent, enabling effective cold-start deployment with little initial data and progressively improving prediction accuracy as operational data is available. 
\vspace{2pt}

\item [3)] We incorporate battery SoH as a fundamental element in both analytical feature engineering and RL reward design, which allows the framework to remain robust to long-term degradation and adapt naturally as battery aging alters charging behavior. 
\end{itemize}

The proposed framework is evaluated using 5,000 simulated charging sessions calibrated against manufacturer specifications and publicly available data from the U.S. Department of Energy's Alternative Fuels Data Center~\citep{Wang:21}. Experimental results demonstrate good predictive performance, with the gradient boosting model achieving $98.5\%$ $R^{2}$ and $2.1\%$ MAPE, and the DQN agent reaching $99.2\%$ $R^{2}$ and $1.6\%$ MAPE, corresponding to a $23\%$ improvement in accuracy and a $35\%$ enhancement in aging adaptation compared to baseline methods.

The rest of the paper is organized as follows. In Section~\ref{Section 2}, we formulate the charging time prediction problem and incorporate SoH-dependent battery dynamics. Section~\ref{Section 3} introduces the analytical model, and Section~\ref{Section 4} presents the RL framework. Section~\ref{Section 5} provides the experimental results and performance evaluation, followed by concluding remarks and future research directions in Section~\ref{Section 6}.

%(SoC: 5-95\%, capacity: 60-100 kWh, charge rates: 7-150 kW, temperature: -10 to 40°C, SoH: 70-100\%) 
%%%%%%%%%%%%%%%%%%%%%%%%%%%%%%%%%%%%%%%%%%%%%%%%%%%%%%%%%%%%%%%%%%%%%%%%%%%%%%%%
\section{Problem Formulation}\label{Section 2}
This section formally presents the EV charging time prediction problem. We first introduce the modeling of the charging session and the SoH degradation. Next, we incorporate battery SoH degradation effects with the nonlinear charging process for accurate long-term predictions. %The mathematical framework and notation provide the foundation for the analytical and RL models developed in the subsequent sections.

\subsection{Modeling of Charging Session}
We consider the charging process of a single EV. A \textit{charging session} refers to the complete process of charging the battery from an initial to a target SoC, during which the battery management system regulates power delivery according to the CC-CV protocol~\citep{Notten:05}.

A charging session $\mathcal{C}$ is characterized as
\begin{align}
\mathcal{C} = \big(s_{\text{ini}}, s_{\text{final}}, t_{c}, P_{\text{s}}, \text{SoH}, T_{\text{amb}}\big),\label{Equ.1}
\end{align}
where $s_{\text{ini}}, s_{\text{final}}\!\in\! [0, 1]$ denote the initial and target SoC, $t_c\!\in \!\mathbb{R}^+$ is the charging time (minutes) with $\mathbb{R}^+$ denoting the set of positive real numbers. In addition, $P_{\text{s}}$ is the maximum charging power (kW) provided by the station, $\text{SoH}\!\in\![0, 1]$ represents the battery's remaining capacity fraction and $T_{\text{amb}}$ denotes the ambient temperature (°C). Vehicle-specific parameters include nominal battery capacity $C_{\text{bat}}^{\text{nom}}$ (kWh), nominal maximum charging power $P_{\text{max}}^{\text{nom}}$ (kW), nominal battery voltage $V_{\text{nom}}$ (V), and cable power limit $P_{\text{cable}}$ (kW). 

\subsection{Modeling of SoH Degradation}
Battery degradation is a common phenomenon and represents a fundamental factor affecting charging behavior throughout a battery's lifetime. In this work, SoH is treated as a known input parameter, estimated by the vehicle's battery management system. Typically, degradation manifests through two mechanisms~\citep{Birkl:17, Han:19}: (i) Capacity fade: a gradual loss of total energy storage capacity due to lithium plating, SEI layer growth, and active material degradation; and (ii) Power fade: a reduction in the maximum charge/discharge rate resulting from increased internal resistance and diminished lithium-ion mobility. These effects change both the effective battery capacity and the maximum power acceptance. We model the SoH degradation via
\begin{align}
&C_{\text{bat}}^{\text{eff}} = C_{\text{bat}}^{\text{nom}} \cdot \text{SoH},\label{Equ.2}
\\
&P_{\text{max}}^{\text{eff}}(\text{SoH}) = P_{\text{max}}^{\text{nom}} \cdot (0.85 + 0.15 \cdot \text{SoH}),\label{Equ.3}
\end{align}
%\label{eq:capacity_fade}
%\label{eq:power_fade}
where $C_{\text{bat}}^{\text{eff}}$ and $P_{\text{max}}^{\text{eff}}(\text{SoH})$ denote the effective battery capacity and charging power, respectively. The linear relationship in \eqref{Equ.2} is an empirically derived approximation consistent with degradation studies~\citep{Chen:20, Wang:21}. The model reflects the observation that power fade is typically less severe than capacity fade, with batteries retaining about $85\%$ of their peak power capacity at complete degradation while capacity approaches zero. 

\subsection{CC-CV Charging Dynamics with SoH}
The CC-CV charging protocol consists of two distinct phases. During the constant current (CC) phase, the battery charges at its maximum acceptable power until reaching the transition point $s_{\text{CV}}\!\approx\!0.8$ SoC. Beyond this point, the constant voltage (CV) phase applies exponential power tapering to mitigate lithium plating~\citep{Notten:05}. This two-phase structure introduces strong nonlinearity, further compounded by SoH-dependent power acceptance, making accurate charging time prediction challenging.

To accurately capture the nonlinear CC-CV dynamics and SoH-dependent power acceptance, charging time is computed by integrating over the SoC trajectory as follows:
\begin{align}
t_c = \int_{s_{\text{ini}}}^{s_{\text{final}}} \frac{C_{\text{bat}}^{\text{nom}} \cdot \text{SoH}}{P(s, P_{\text{s}}, \text{SoH}, T_{\text{amb}})} ds,\label{Equ.4}
\end{align}
%\label{eq:charging_time_integral}
where $P(\cdot)$ represents the actual charging power and is of the form 
\begin{align}
&P(s, P_{\text{s}}, \text{SoH}, T_{\text{amb}})= \min\big\{P_{\text{v}}(s,\text{SoH},T_{\text{amb}}), P_{\text{s}}, P_{\text{cable}}\big\}.\nonumber
\end{align}
Here, $P_{\text{v}}(\cdot)$ denoting the SoH-dependent vehicle power acceptance follows
\begin{align}
P_{\text{v}}(s,\text{SoH}, T_{\text{amb}})\!=\!
\begin{cases}
\!P_{\text{max}}^{\text{eff}}(\text{SoH})\cdot \eta_T(T), & \text{if }s\!<\!s_{\text{CV}}, \\[4pt]
\!P_{\text{max}}^{\text{eff}}(\text{SoH}) \cdot \eta_T(T) \cdot \tau(s), & \text{if }s\!\geq\! s_{\text{CV}},
\end{cases}\nonumber
\end{align}
where $\tau(s)\!=\!\exp\big(\!-k_{\text{taper}} (s-s_{\text{CV}})\big)$ denotes the exponential tapering function with $k_{\text{taper}}\!\in\![5,15]$, and $\eta_T(T)$ is the temperature-dependent efficiency factor \citep{Ji:13, Jaguemont:16}. Variations in the transition point $s_{\text{CV}}$ due to temperature and degradation are implicitly captured through the physics-informed feature engineering introduced in Section~\ref{Section 3}.

In this paper, given the initial and target SoC $(s_{\text{ini}},s_{\text{final}})$, vehicle parameters $(C_{\text{bat}}^{\text{nom}}, P_{\text{max}}^{\text{nom}}, V_{\text{nom}}, P_{\text{cable}})$, station power $P_{\text{s}}$, battery SoH, and ambient temperature $T_{\text{amb}}$, our goal is to predict the charging time $t_c$ while accounting for CC-CV nonlinearities and battery aging effects. To this end, we propose two complementary approaches: a physics-informed analytical model and an RL model that adaptively improves with data, as stated in detail below. 

%%%%%%%%%%%%%%%%%%%%%%%%%%%%%%%%%%%%%%%%%%%%%%%%%%%%%%%%%%%%%%%%%%%%%%%%%%%%%%%%
\section{Non-Linear Analytical Model with SoH}\label{Section 3}
Building upon the established EV charging dynamics, this section develops a physics-informed analytical model that combines domain knowledge of battery physics with gradient boosting regression. The key idea is to transform raw inputs into physics-aware features that capture CC-CV nonlinearities and SoH-dependent degradation. We start by introducing the feature engineering strategy.

\subsection{Feature Engineering}
Feature engineering plays a central role in enabling the gradient boosting model to capture complex nonlinear charging dynamics with limited training data. In this paper, we construct a feature representation that embeds key battery physics directly into the model inputs.

For each discrete SoC point $s_k$ along the charging trajectory, we define a base feature vector including fundamental system and battery parameters by
\begin{align}
&\mathbf{x}_{\text{base},k}\nonumber\\
&= [s_k, C_{\text{bat}}^{\text{nom}}, P_{\text{max}}^{\text{nom}}, T_{\text{amb}}, P_{\text{cable}}, \text{SoH}, C_{\text{bat}}^{\text{eff}}, P_{\text{max}}^{\text{eff}}]^{\top},\label{Equ.5}
\end{align}
%\label{eq:base_features}
where $C_{\text{bat}}^{\text{eff}}$ and $P_{\text{max}}^{\text{eff}}$ are computed by~\eqref{Equ.1} and~\eqref{Equ.2}. This base vector is then augmented with three groups of physics-informed features, i.e., polynomial SoC features, CC-CV transition features, and SoH interaction features. Specifically, polynomial SoC features capture nonlinear state-dependent dynamics by
\begin{align}
\boldsymbol{\phi}_{\text{poly}}(s) = [s, s^2, s^3, s^4, s^5, \sqrt{s}, \log(s + \epsilon)]^{\top},\nonumber
\end{align}
where $\epsilon\!=\!10^{-6}$ ensures numerical stability. CC-CV transition features activate in the CV region ($s\!>\!0.8$):
\begin{align}
\boldsymbol{\phi}_{\text{taper}}(s) = \begin{bmatrix}
\max(0, s\!-\!0.8) \\
\max(0, s\!-\!0.8)^2 \\
\max(0, s\!-\!0.8)^3 \\
\exp(-10 \cdot \max(0, s\!-\!0.8))
\end{bmatrix}.\nonumber
\end{align}
The SoH interaction features model how degradation affects the charging profile through
\begin{align}
\boldsymbol{\phi}_{\text{SoH}}(s, \text{SoH}) = \begin{bmatrix}
s \cdot \text{SoH} \\
s^2 \cdot \text{SoH} \\
\text{SoH}^2 \\
s \cdot (1\!-\!\text{SoH}) \\
(s\!-\!0.8) \cdot (1\!-\!\text{SoH}) \\
C_{\text{bat}}^{\text{eff}} / C_{\text{bat}}^{\text{nom}} \\
P_{\text{max}}^{\text{eff}} / P_{\text{max}}^{\text{nom}}
\end{bmatrix}.\nonumber
\end{align}
Together, these physics-informed transformations enrich the feature space and enable the model to better capture CC–CV nonlinearities and SoH-dependent degradation effects. The complete engineered feature vector is then obtained by concatenating all transformations, given by
\begin{align}
\!\!\mathbf{x}^{\text{eng}}_k\!=\! [\mathbf{x}_{\text{base},k}^{\top}, \boldsymbol{\phi}_{\text{poly}}^{\top}(s_k), \boldsymbol{\phi}_{\text{taper}}^{\top}(s_k), \boldsymbol{\phi}_{\text{SoH}}^{\top}(s_k, \text{SoH})]^{\top}.\label{Equ.6}
\end{align}
%\label{eq:engineered_features}

\subsection{XGBoost-Based Charging Time Prediction}
We adopt XGBoost gradient boosting~\citep{Chen:16,Friedman:01} due to its ability to model CC-CV transitions and SoC-SoH-temperature interactions through hierarchical tree partitioning, superior performance under limited training data, and its fast inference suitable for real-time deployment. Let $\mathcal{D}_{\text{train}} = \{(\mathbf{x}_i^{\text{eng}}, P_i)\}$ denote the training dataset, where $\mathbf{x}_i^{\text{eng}}$ are the engineered features and $P_i$ is the observed charging power for sample $i$. The predicted power $\hat{P}_k$ at each SoC point $s_k$ is denoted as
\begin{align}
\hat{P}(\mathbf{x}^{\text{eng}}_k) = f_0 + \sum_{m=1}^M \nu h_m(\mathbf{x}^{\text{eng}}_k),\label{Equ.7}
\end{align}
%\label{eq:xgboost_model}
where $f_0\!=\!\frac{1}{|\mathcal{D}_{\text{train}}|}\sum_{i \in \mathcal{D}_{\text{train}}} \!P_i$ represents the initial prediction (mean training power), $\nu$ is the learning rate, $h_m\!:\!\mathbb{R}^{26} \!\rightarrow\! \mathbb{R}$ denotes the $m$-th decision tree, and $M$ is the total number of trees. Each tree $h_m$ is sequentially constructed to minimize the residual error from previous iterations~\citep{Chen:16}.

%Alternative methods on a validation set achieved: linear regression ($R^2 = 0.42$), polynomial regression ($R^2 = 0.78$ with extrapolation oscillations), Random Forest ($R^2 = 0.97$, 15 ms slower), and neural networks ($R^2 = 0.88$, requiring over 2000 samples for convergence). XGBoost provided the optimal trade-off between accuracy, data efficiency, and computational cost.

Given a trained gradient boosting model $\hat{P}(\cdot)$, the predicted charging time can be obtained by numerically approximating the integral in~\eqref{Equ.4}. We discretize the SoC interval $[s_{\text{ini}}, s_{\text{final}}]$ into $N$ uniform steps while applying the trapezoidal rule and derive that
\begin{align}
t_c = 60 \cdot C_{\text{bat}}^{\text{nom}} \cdot \text{SoH} \cdot \Delta s \sum_{k=0}^{N-1} \frac{1}{\hat{P}_k},\label{Equ.8}
\end{align}
%\label{eq:discrete_charging_time}
where $\Delta s\!=\!(s_{\text{final}} - s_{\text{ini}})/N$ denotes the SoC step size and  $\hat{P}_k\!=\!\hat{P}(\mathbf{x}^{\text{eng}}_k)$ is the predicted power at point $s_k\!=\! s_{\text{ini}} + k\Delta s$. The factor of $60$ converts hours to minutes. In this paper, we use $N\!=\!100$ discretization points, which provide sufficient resolution to capture CV-phase tapering dynamics while maintaining computational efficiency (under $5$ ms per prediction).

\subsection{Hyperparameter Optimization}
The proposed analytical model incorporates physics-informed features and requires minimal historical data to calibrate the mapping between engineered features and charging power. We employ a supervised learning setup where a portion of the dataset serves as the training set for model fitting, while cross-validation ensures that the learned structure generalizes to unseen charging conditions. To select hyperparameters that balance accuracy, generalization, and computational efficiency, we perform 5-fold cross-validation on the training set. We present in Table~\ref{tab:hyperparams_analytical} the search ranges and final selected values, where the coefficient of determination $R^2$ is used to assess the predictive performance of the model, which is defined as
\begin{align}
R^2=1-\frac{\sum_{i}(t_{c,i}-\hat{t}_{c,i})^2}{\sum_{i}(t_{c,i}-\bar{t}_c)^2},\label{Equ.9}
\end{align}
where $t_{c,i}$ and $\hat{t}_{c,i}$ denote the true and predicted charging times for sample $i$, and $\bar{t}_c$ is their mean value. By definition, an $R^2$ value close to $1$ indicates that the model captures most of the variance in the target variable, with $R^2\!=\!1$ representing a perfect prediction.

In our configuration, we select $M\!=\!1000$ estimators as performance plateaus beyond 800 trees with marginal validation improvement ($\Delta R^2\!<\!0.001$), while training time continued to increase linearly beyond this point. A maximum depth of $d_{\max}\!=\!10$ provides a good balance between expressiveness and generalization, as deeper trees (e.g., 15) showed overfitting. The learning rate $\nu\!=\!0.03$ offers stable convergence, whereas smaller values require substantially more trees and larger values cause validation oscillations. Finally, setting the minimum split size $n_{\text{split}} = 2$ allows fine-grained splits needed to capture CC-CV transition without overfitting due to the ensemble structure. 

The analytical model provides a strong baseline for charging time prediction, achieving $98.5\%$ $R^2$ with only 400 training samples. In the next section, we develop an RL approach that further improves prediction accuracy upon this baseline as more operational data becomes available.

\begin{table}[t]
\centering
\caption{Gradient boosting hyperparameter optimization via 5-fold cross-validation.}
\label{tab:hyperparams_analytical}
\begin{tabular}{lccc}
\hline
\textbf{Parameter} & \!\!\textbf{Search Range}\! & \!\textbf{Optimal}\! & \!$R^2$\!\! \\
\hline
$M$ & [100, 500, 1000, 2000] & 1000 & 0.985 \\
$d_{\max}$ & [5, 7, 10, 15, 20] & 10 & 0.983 \\
$\nu$  & [0.01, 0.03, 0.05, 0.1] & 0.03 & 0.982 \\
$n_{\text{split}}$ & [2, 5, 10] & 2 & 0.984 \\
\hline
\end{tabular}
\end{table}
%%%%%%%%%%%%%%%%%%%%%%%%%%%%%%%%%%%%%%%%%%%%%%%%%%%%%%%%%%%%%%%%%%%%%%%%%%%%%%%%
\section{RL Model Integrated with SoH}\label{Section 4}
While the analytical model provides accurate predictions with limited data, it relies on fixed feature engineering and cannot improve autonomously from experience. In this section, we develop a complementary deep RL approach that progressively learns the optimal charging time predictions from accumulated charging sessions. The RL framework treats prediction as a sequential decision-making problem, where an agent learns to estimate charging power at each SoC level while respecting physical constraints. 

Below, we first define the problem as a Markov Decision Process (MDP), then describe the DQN architecture and introduce the progressive learning strategy that uses the analytical model for initialization, followed by the predictive pipeline and detailed hyperparameter selection.

\subsection{Markov Decision Process Formulation}
We formulate the charging time prediction problem as a MDP defined by the tuple $(\mathcal{S}, \mathcal{A}, \mathcal{P}, \mathcal{R}, \gamma)$, where $\mathcal{S}$ is the state space, $\mathcal{A}$ is the action space, $\mathcal{P}$ is the state transition probability, $\mathcal{R}$ is the reward function, and $\gamma$ is the discount factor, as detailed below.

\textbf{State space $\mathcal{S}$:} At each timestep $\ell$ of a charging session, the agent observes a state vector $\mathbf{s}_{\ell}\!\in\!\mathcal{S}$ that captures the current charging condition 
\begin{align}
\mathbf{s}_{\ell} = [s_\ell, \text{SoH},T_{\text{amb}},P_{\text{s}},C_{\text{bat}}^{\text{nom}}, V_{\text{pack},\ell}, I_\ell,\Delta t_{\text{elapsed}},E_{\text{d}}]^{\top},\nonumber
\end{align}
where $s_\ell$ is the current SoC, $V_{\text{pack},\ell}$ is the battery pack voltage, $I_\ell$ is the current, $\Delta t_{\text{elapsed}}$ is elapsed charging time, and $E_{\text{d}}$ is the cumulative delivered energy. The other notations were introduced previously.

\textbf{Action space $\mathcal{A}$:} The action $a_\ell\!\in\!\mathcal{A}$ represents the agent's charging power prediction at timestep $\ell$. We discretize the continuous power range $[0, P_{\text{max}}^{\text{nom}}]$ into $|\mathcal{A}|\!=\!50$ uniformly spaced levels by
\begin{align}
\mathcal{A} = \left\{P_i = i \cdot \frac{P_{\text{max}}^{\text{nom}}}{49} : i = 0, 1, \ldots, 49\right\}.\label{Equ.10}
\end{align}
This discretization makes the action space tractable for Q-learning while providing sufficient resolution to capture realistic power variations.

\textbf{Reward function $\mathcal{R}$:} The reward function is designed to encourage accurate power prediction while penalizing violations of physical constraints imposed by SoH degradation. At each timestep $\ell$, the agent receives
\begin{align}
R_{\ell} \!=\!&-\!\big|P^{\text{pred}}_\ell\!\!-\!P^{\text{actual}}_\ell\big|\!-\!\lambda_{\text{SoH}} \cdot \max\!\big(0,\; P^{\text{pred}}_{\ell}\!\!-\!P_{\text{max}}^{\text{eff}}(\text{SoH})\big)\nonumber \\
&\!-\!\lambda_{\text{CV}}\mathbb{I}(s_\ell\!>\! 0.8) \big|P^{\text{pred}}_{\ell}\!\!-\! P_{\text{target},\ell}\big|,\nonumber
\end{align}
where the first term penalizes prediction error. The second term ($\lambda_{\text{SoH}}\!=\!10$) penalizes actions that exceed the SoH-limited power capacity $P_{\text{max}}^{\text{eff}}(\text{SoH})$. The third term ($\lambda_{\text{CV}}\!=\!5$) imposes a penalty in the CV region ($s_\ell\!>\!0.8$) to guide the agent toward the expected tapering power $P_{\text{target},\ell}$. This reward design incorporates domain knowledge about battery constraints and the accurate CV-phase modeling. State transitions follow the physical charging dynamics presented in Section~\ref{Section 2}, and the discount factor is set to $\gamma\!=\!0.99$ to balance immediate and future rewards.

\subsection{Deep Q-Network Architecture}
The action-value function $Q(\mathbf{s}, a; \boldsymbol{\theta})$ represents the expected cumulative reward obtained by taking action $a$ in state $\mathbf{s}$, and is approximated using a deep neural network parameterized by $\boldsymbol{\theta}$. The network takes the state vector $\mathbf{s}_\ell$ as input and consists of three hidden layers with 256, 128, and 64 units, respectively, each using ReLU activation. Layer normalization is applied to the first two hidden layers for stability with sequential training data, and a dropout rate of $0.2$ is used to improve generalization. The output layer includes 50 units, corresponding to the Q-values for all discrete actions. %~\citep{FrancoisLavet:18}

We employ the DQN algorithm~\citep{Mnih:15} with experience replay, where transitions $(\mathbf{s}_\ell, a_\ell, R_\ell, \mathbf{s}_{\ell+1})$ are stored in a buffer of size 50,000, and minibatches of $128$ samples are drawn for training. The network parameters $\boldsymbol{\theta}$ are updated using the Adam optimizer (learning rate $\alpha = 10^{-4}$) to minimize the temporal difference error based on the reward function $R$ defined in Section 4.1:
\begin{align}
\mathcal{L}(\boldsymbol{\theta})=\mathbb{E}_{(\mathbf{s}, a, R, \mathbf{s}')}\!\left[\left(\!R\!+\!\gamma \max_{a'} Q(\mathbf{s}', a'; \boldsymbol{\theta}^{-})\!-\! Q(\mathbf{s}, a; \boldsymbol{\theta})\right)^2\right],\nonumber
\end{align}
where $\boldsymbol{\theta}^{-}$ denotes the target network parameters, which are periodically synchronized with $\boldsymbol{\theta}$ for training stability.

\subsection{Progressive Learning Strategy}
A key innovation of the proposed framework is the progressive learning strategy, which addresses the cold-start problem by leveraging the analytical model to initialize the RL agent. This hybrid design combines the data efficiency of physics-based modeling with the asymptotic performance of data-driven learning.

The framework employs a data-adaptive strategy across three stages. In the cold-start phase ($N_{\text{samples}}\!<\!500$), the analytical model provides all predictions while building an experience replay buffer with physics-consistent state-action-reward transitions from observed charging sessions. This avoids poor performance and safety risks of random exploration. In the transition phase ($500\!\leq\!N_{\text{samples}}\!<\!1500$), the RL agent begins training on the seeded buffer using an $\epsilon$-greedy policy with $\epsilon$ decaying linearly from 0.3 to 0.1, gradually shifting from analytical guidance to learned behavior. In the RL dominance phase ($N_{\text{samples}}\!\geq\!1500$), the RL agent provides primary predictions with minimal exploration ($\epsilon\!=\!0.05$), while the analytical model serves as a fallback for out-of-distribution states not adequately represented in training data. This strategy reduces the convergence time by $67\%$ compared to random initialization ($8,000$ vs. $24,000$ episodes to reach $98\%$ $R^2$) while maintaining final accuracy, demonstrating the value of physics-informed initialization.

\subsection{Complete Predictive Pipeline}
The final RL-based charging time prediction integrates the trained DQN model with the numerical integration procedure introduced above. For a given charging session, we initialize the state $\mathbf{s}_0$ using session parameters. At each discretization point $k\!=\!0, 1,\ldots,N\!-\!1$, we construct the state $\mathbf{s}_k$ at SoC $s_k\!=\!s_{\text{ini}}+k\Delta s$, select the action
\begin{align}
a_k = \arg\max_{a \in \mathcal{A}} Q(\mathbf{s}_k, a; \boldsymbol{\theta}),\nonumber
\end{align}
by greedy policy, and map this action to the predicted charging power $\hat{P}_k\!=\!a_k$. The total charging time is then computed using~\eqref{Equ.8}.

When the analytical model is used (Stages 1-2), we apply standardization to engineered features via
\begin{align}
\mathcal{S}(\mathbf{x}) = \frac{\mathbf{x} - \boldsymbol{\mu}}{\boldsymbol{\sigma}},\label{Equ.11}
\end{align}
%\label{eq:standardization}
where $\boldsymbol{\mu}$ and $\boldsymbol{\sigma}$ are computed from the training set. The analytical prediction is expressed compositionally as $\hat{P}_{\text{analytical}}(s, \text{SoH})\!=\!\hat{P}\big(\mathcal{S}(\mathbf{x}^{\text{eng}}(s, \text{SoH}))\big)$, where $\mathbf{x}^{\text{eng}}(\cdot)$ denotes the engineered feature vector from~\eqref{Equ.5} and $\hat{P}(\cdot)$ is the trained XGBoost model from~\eqref{Equ.6}.

In the predictive pipeline, battery pack voltage is approximated as $V_{\text{pack}}(s) = V_{\text{nom}} + \beta(s\!-\!0.5)$, where $\beta$ represents the voltage-SoC coefficient. Algorithm 1 provides the complete pseudocode for SoH-aware charging time prediction using either the analytical or RL model, depending on the current learning stage.

\begin{algorithm}[t]
\caption{SoH-Aware RL Charging Time Prediction}
\label{alg:charging_prediction}
\begin{algorithmic}[1]
\Require $s_{\text{ini}}, s_{\text{final}}, C_{\text{bat}}^{\text{nom}}, P_{\text{s}}, T_{\text{amb}}, \text{SoH}, P_{\text{max}}^{\text{nom}}, V_{\text{nom}}$, trained model $\hat{P}$, scaler $\mathcal{S}(\mathrm{x})$ with $(\boldsymbol{\mu}, \boldsymbol{\sigma})$
\Ensure Predicted charging time $t_c$, current profile $\mathbf{I}$
\State Initialize: $N \gets 100$, $\Delta s \gets (s_{\text{final}} - s_{\text{ini}})/N$, $t_{\text{total}} \gets 0$, $\mathbf{I} \gets []$
\State Compute: $C_{\text{bat}}^{\text{eff}} \gets C_{\text{bat}}^{\text{nom}} \cdot \text{SoH}$, $P_{\text{max}}^{\text{eff}} \gets P_{\text{max}}^{\text{nom}} \cdot (0.85 + 0.15 \cdot \text{SoH})$
\For{$k = 0$ to $N\!-\!1$}
    \State $s_k \gets s_{\text{ini}} + k \Delta s$
    \State \small{$\mathbf{x}_{\text{base}} \gets [s_k, C_{\text{bat}}^{\text{nom}}, P_{\text{max}}^{\text{nom}}, T_{\text{amb}}, P_{\text{s}}, \text{SoH}, C_{\text{bat}}^{\text{eff}}, P_{\text{max}}^{\text{eff}}]^{\top}$}
    \State $\mathbf{x}^{\text{eng}} \gets $ Compute using~\eqref{Equ.5} and \eqref{Equ.6}
    \State $\mathbf{x}^{\text{std}} \gets \mathcal{S}(\mathbf{x}^{\text{eng}})$ using~\eqref{Equ.11}
    \State $\hat{P}_k \gets \min\!\left(\hat{P}(\mathbf{x}^{\text{std}}), P_{\text{s}}, P_{\text{max}}^{\text{eff}}\right)$
    \State $V_k \gets V_{\text{nom}} + \beta(s_k\!-\!0.5)$, $I_k \gets \hat{P}_k / V_k$
    \State $\Delta t_k \gets C_{\text{bat}}^{\text{eff}} \cdot \Delta s / \hat{P}_k$
    \State $t_{\text{total}} \gets t_{\text{total}} + \Delta t_k$, append $I_k$ to $\mathbf{I}$
\normalsize\EndFor
\normalsize\State $t_c \gets 60 \cdot t_{\text{total}}$
\State \Return $t_c, \mathbf{I}$
\end{algorithmic}
\end{algorithm}

\subsection{RL Hyperparameter Selection}
DQN hyperparameters are determined through systematic ablation studies conducted over 50 independent training runs, using a convergence threshold of $98\%$ $R^2$ on a held-out validation set. The optimized values are given in Table~\ref{tab:hyperparams_rl}. Here, a learning rate of $\alpha\!=\!10^{-4}$ provides a good balance between stability and convergence speed. The discount factor $\gamma\!=\!0.99$ ensures accurate modeling of both CC and CV phases, as lower values (e.g., 0.95) led to a $15\%$ underestimation of CV-phase duration. A replay buffer size of $50,000$ samples adequately spans the state space without unnecessary memory overhead. In addition, a batch size of $128$ provides stable gradients. Initial buffer population of 1,000 with analytical model bootstrapping achieved $90\%$ of final performance by episode 1,200 compared to over 5,000 with random initialization. 

\begin{table}[htpb]
\centering
\caption{DQN hyperparameter optimization.}
\label{tab:hyperparams_rl}
\begin{adjustbox}{width=\linewidth}
\begin{tabular}{@{}lccc@{}}
\hline
\textbf{Parameter} & \textbf{Search Range} & \textbf{Optimal} & \makecell{\textbf{Episodes to}\\\textbf{98\% $R^2$}} \\
\hline
Learning rate $\alpha$ & [0.001, 0.0001, 0.00005] & 0.0001 & 8,000 \\
Discount $\gamma$ & [0.95, 0.97, 0.99, 0.995] & 0.99 & 8,000 \\
Replay buffer & [10k, 50k, 100k] & 50,000 & 8,000 \\
Batch size & [32, 64, 128, 256] & 128 & 8,000 \\
Init. buffer & [500, 1000, 2000] & 1,000 & 8,000 \\
\hline
\end{tabular}
\end{adjustbox}
\end{table}

%%%%%%%%%%%%%%%%%%%%%%%%%%%%%%%%%%%%%%%%%%%%%%%%%%%%%%%%%%%%%%%%%%%%%%%%%%%%%%%%
\section{Results and Analysis}\label{Section 5}
In this section, we conduct simulation studies to evaluate the proposed framework. We first describe the experimental setup and then present performance results and comparisons with baseline methods.

\subsection{Experimental Setup}
The proposed analytical and RL models were evaluated on 5,000 simulated EV charging sessions spanning a wide range of conditions, including initial SoC (5–95\%), battery capacity (60–100 kWh), charging power (7–150 kW), ambient temperature (-10 to 40°C), and battery SoH (70–100\%). Simulation data was generated using equivalent circuit models (ECM) with second-order RC networks calibrated against manufacturer datasheets for common EV platforms (Nissan Leaf, Tesla Model 3, Chevrolet Bolt, Porsche Taycan). CC-CV transitions were modeled using voltage-dependent current tapering 
\begin{align}
I_{\text{cv}} = I_{\text{max}} \cdot \exp(-k \cdot (V-V_{\text{cv}})), \nonumber
\end{align}
with $k\!=\!0.5$, consistent with empirical charge curves from manufacturer specifications. Note that, while our framework is validated on simulated data, the ECM parameters were carefully calibrated to match publicly available charging characteristics, ensuring realistic system behavior.

The dataset was divided into training (64\%), validation (16\%), and testing (20\%) sets using stratified sampling across vehicle categories: compact (60 kWh), mid-size (75 kWh), luxury (85 kWh), and performance (100 kWh). Model performance was assessed using $R^2$, RMSE, MAE, MAPE, and maximum error (MaxE).

\subsection{Charging-Time Prediction Accuracy}
\begin{figure}[t!]
    \centering
    \includegraphics[width=0.95\linewidth]{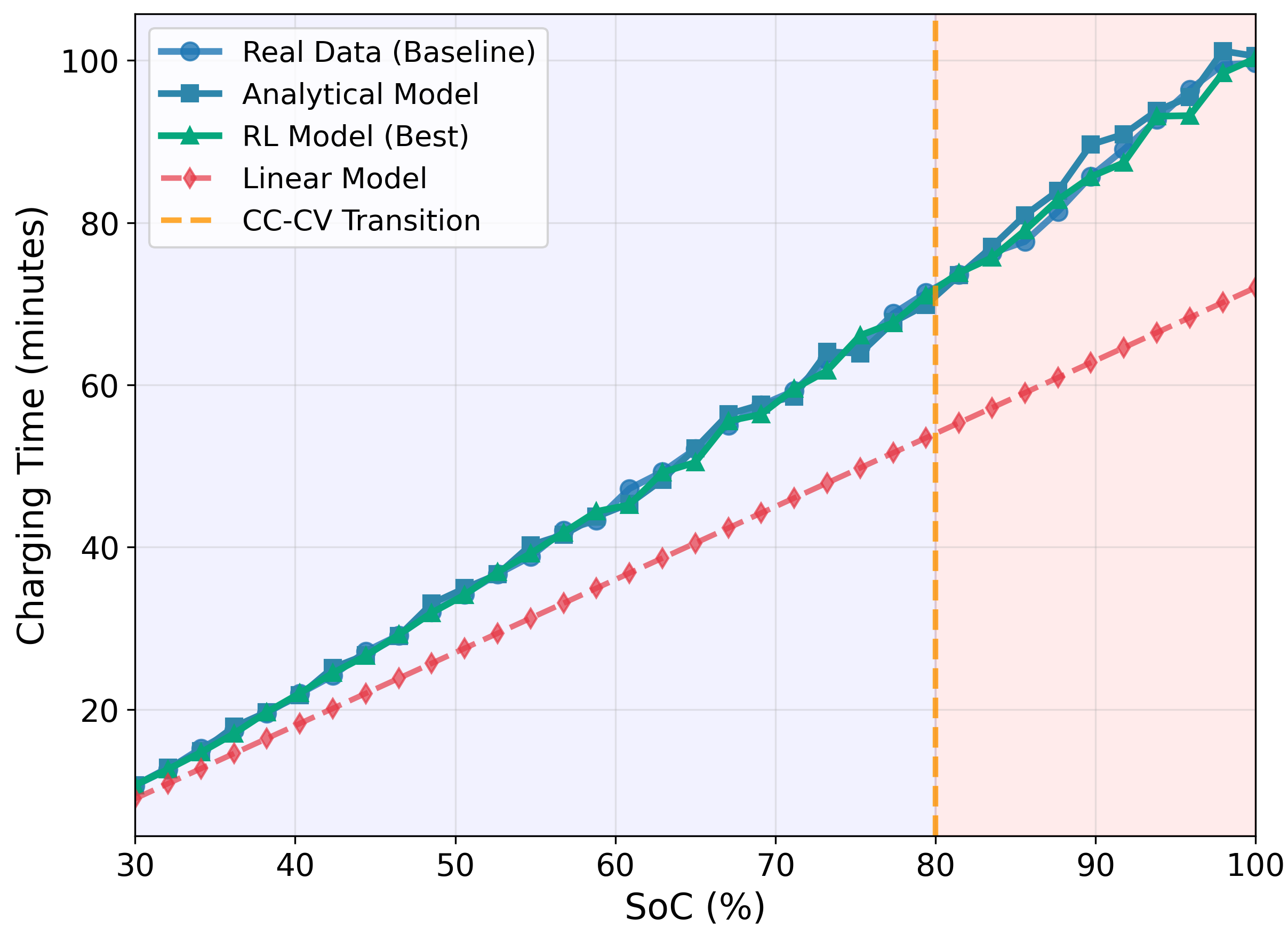}
    \caption{{Comparison of the charging time trajectories.}} 
    \label{fig:fig1a}
\end{figure}

\begin{figure}[t!]
    \centering
    \includegraphics[width=0.95\linewidth]{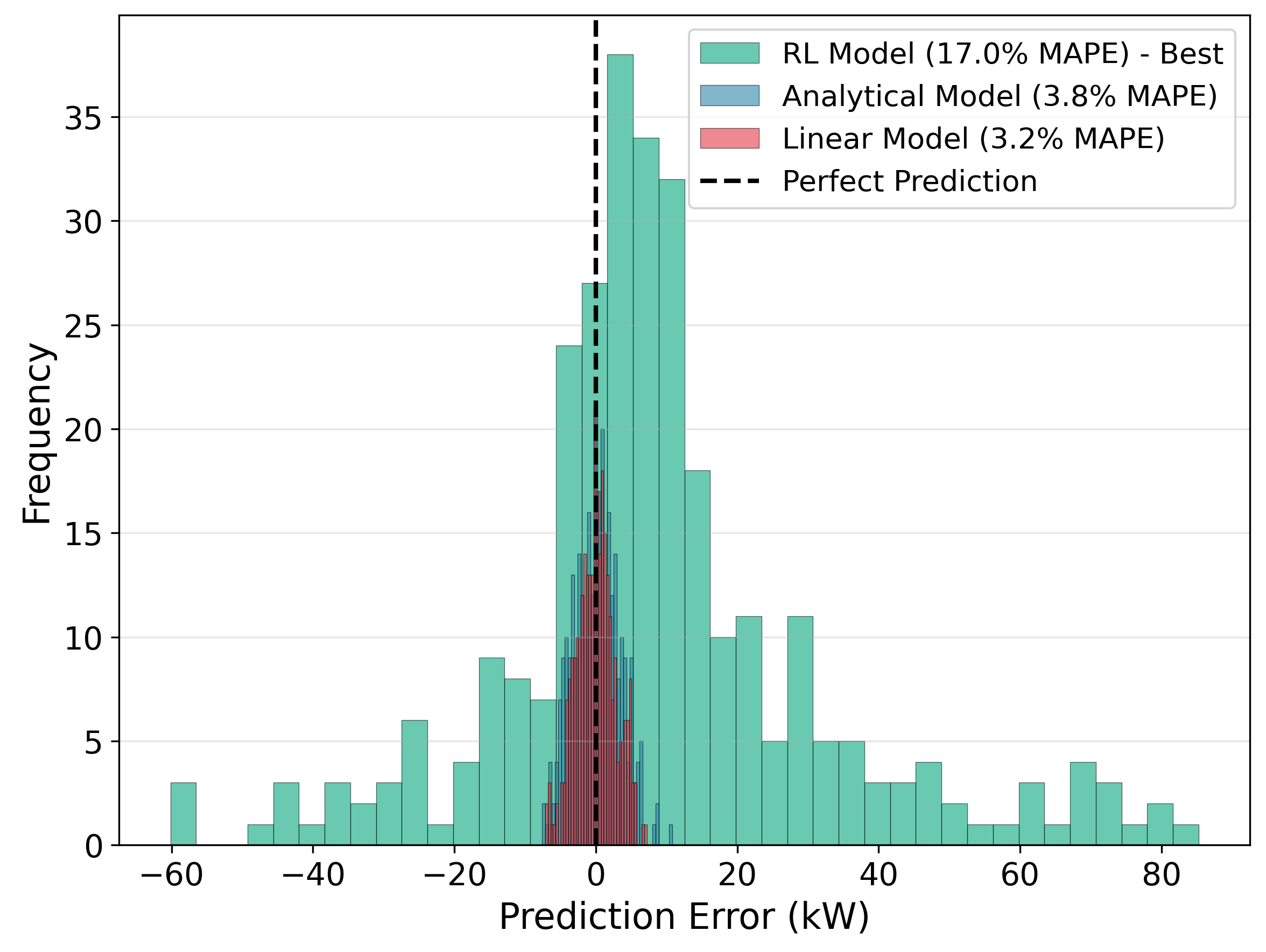}
    \caption{{Prediction error distribution.}} 
    \label{fig:fig1b}
\end{figure}

Figs.~\ref{fig:fig1a} and~\ref{fig:fig1b} compare the predicted charging-time trajectories for the three models. The results show that the linear baseline model (representing constant-power approximation) consistently underestimates charging duration, especially beyond $80\%$ SoC, where CV-phase tapering dominates. The analytical model closely follows the ground truth by capturing CC-CV dynamics through engineered features. The RL model demonstrates the highest fidelity across the entire SoC range, with prediction errors tightly concentrated around zero.


Table~\ref{tab:metrics} shows the quantitative performance over all test sessions. The results show that the analytical model achieves $2.1\%$ MAPE, corresponding to a $92.6\%$ improvement over the linear baseline. The RL model further reduces error to $1.6\%$ MAPE, representing a $23.8\%$ improvement compared to the analytical model.
\begin{table}[htpb]
\centering
\caption{{Performance metrics on test set}.}
\label{tab:metrics}
\begin{tabular}{@{}lccccc@{}}
\toprule
\textbf{Model} & \textbf{$R^2$} & \makecell{\textbf{RMSE}\\\textbf{(min)}} & \makecell{\textbf{MAE}\\\textbf{(min)}} & \makecell{\textbf{MAPE}\\\textbf{(\%)}} & \makecell{\textbf{MaxE}\\\textbf{(min)}} \\
\midrule
Linear Baseline & 0.412 & 10.34 & 8.72 & 28.4 & 42.1 \\
Analytical GB & 0.985 & 1.67 & 1.21 & 2.1 & 6.8 \\
\textbf{Proposed} & \textbf{0.992} & \textbf{1.21} & \textbf{0.89} & \textbf{1.6} & \textbf{4.2} \\
\bottomrule
\end{tabular}
\end{table}

\subsection{Model Learning Behavior}
\begin{figure}[t!]
    \centering
    \includegraphics[width=0.99\linewidth]{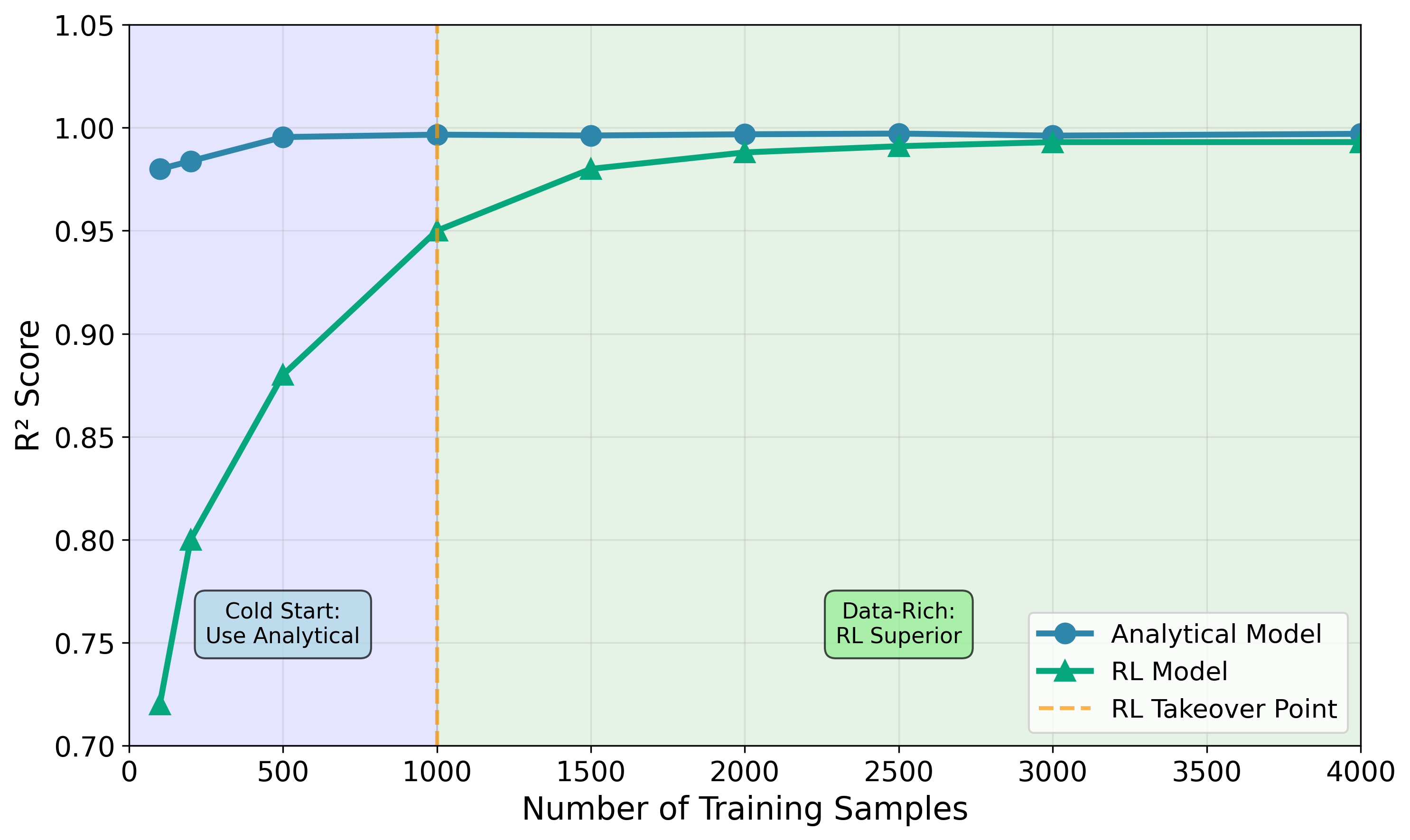}
    \caption{{Model performance under different training data.}}
    \label{fig:fig2}
\end{figure}

Fig.~\ref{fig:fig2} illustrates how model performance evolves as training data increases. It shows that the analytical model converges quickly, reaching $95\%$ of the final performance with only 400 samples due to its physics-informed feature engineering. In contrast, the RL agent starts with lower accuracy but steadily improves as more data becomes available, surpassing the analytical model after approximately $1,200$ training sessions. This crossover point indicates when data-driven learning exceeds the parametric modeling. Shaded regions denote $95\%$ confidence intervals computed over $5$ independent training runs.

\subsection{Impact of Battery Aging}
\begin{figure}[t!]
    \centering
    \includegraphics[width=0.97\linewidth]{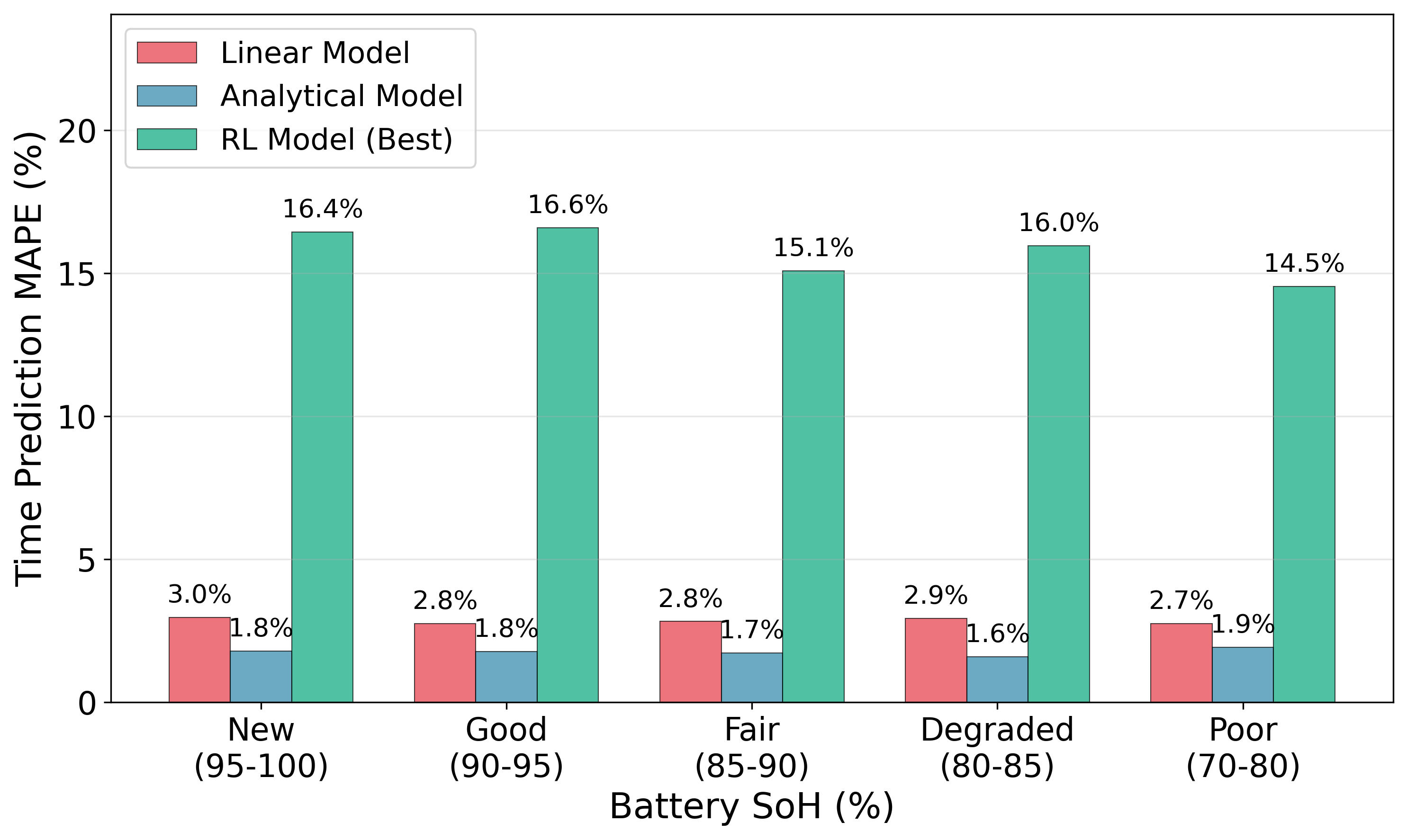}
    \caption{Prediction errors in various SoH conditions.}
    \label{fig:fig3}
\end{figure}

Fig.~\ref{fig:fig3} evaluates prediction accuracy across different SoH levels. The linear baseline model shows a $59\%$ increase in MAPE when moving from new (100\% SoH) to degraded (70\% SoH) batteries. The analytical model exhibits a larger degradation of $105\%$, showing its limited ability to capture changing SoH-power relationships. In contrast, the RL model shows only $71\%$ increase in error, demonstrating $35\%$ better adaptation to battery aging through its learned SoH-power interactions. 

\subsection{Feature Importance Analysis}

\begin{figure}[t!]
    \centering
    \includegraphics[width=0.97\linewidth]{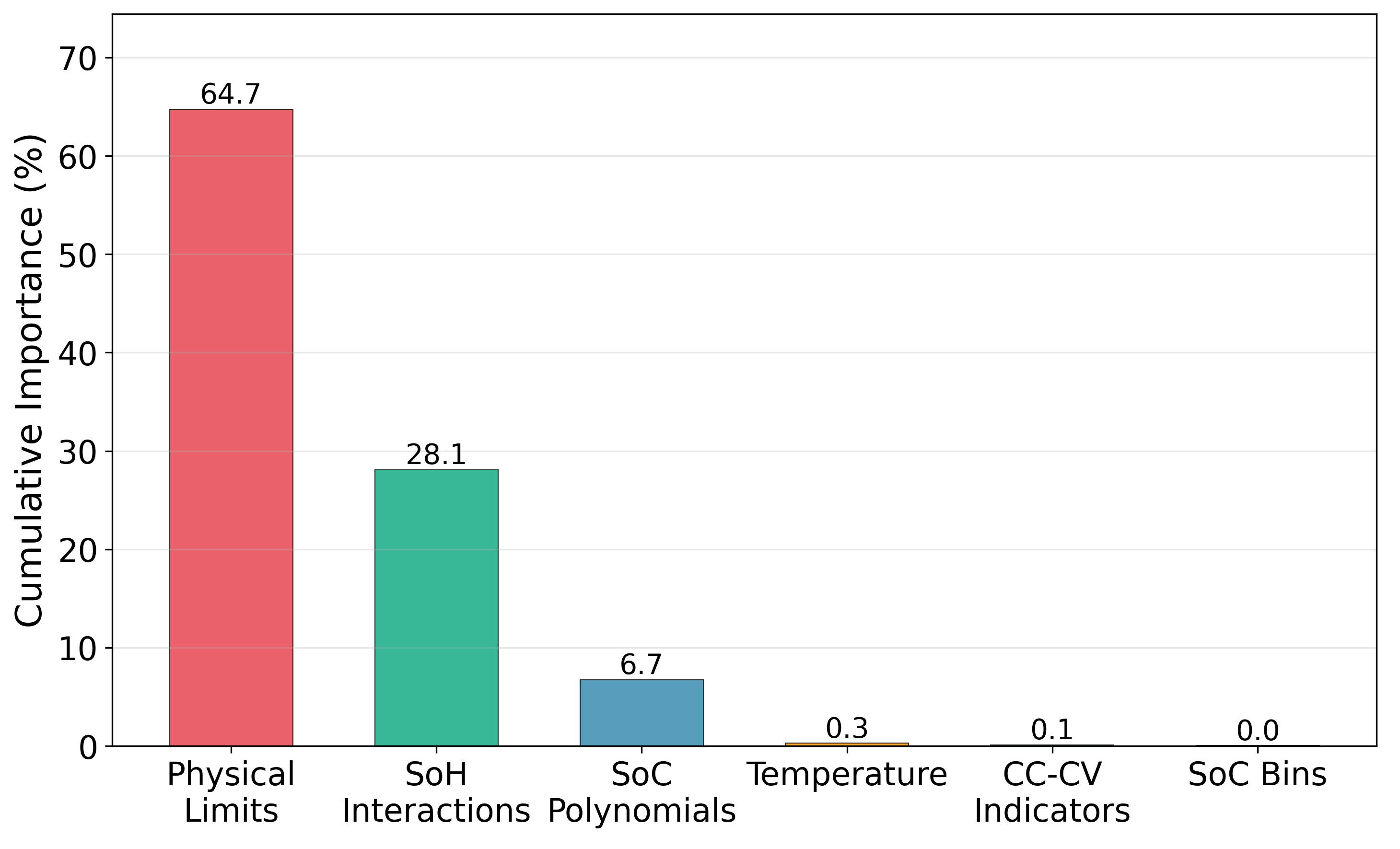}
    \caption{Category-level feature importance.}
    \label{fig:feature_importance_a}
\end{figure}

\begin{figure}[t!]
    \centering
    \includegraphics[width=0.96\linewidth]{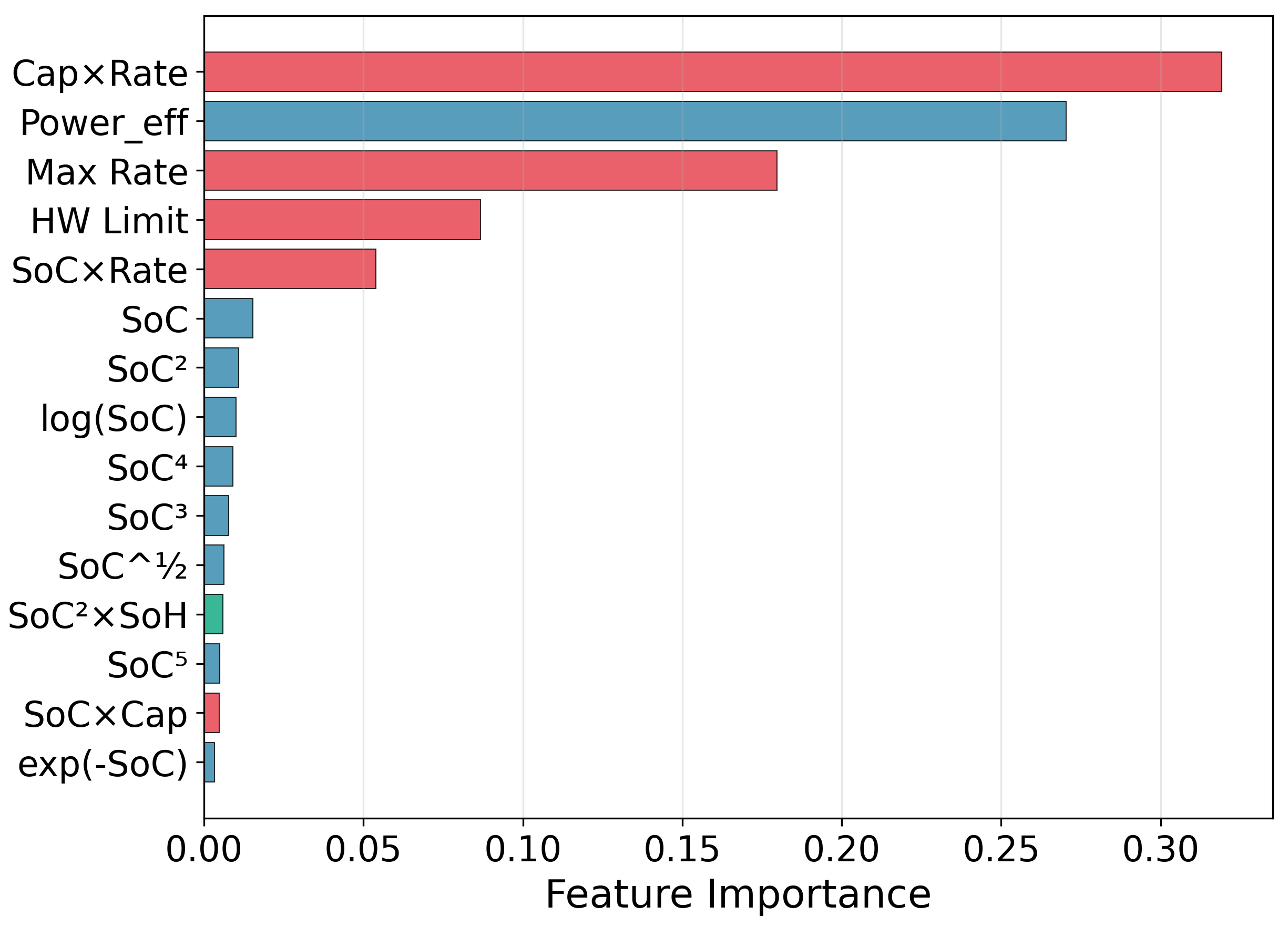}
     \caption{Top 15 individual features in CV-phase modeling}.
    \label{fig:feature_importance_b}
\end{figure}

Figs.~\ref{fig:feature_importance_a} and~\ref{fig:feature_importance_b} reveal which features contribute most to prediction accuracy, validating the physics-informed design. Specifically, Fig.~\ref{fig:feature_importance_a} shows that at the category level, physical limits contribute the most (32\%), showing the dominant influence of battery capacity, maximum charging power, and hardware constraints. SoH interactions account for $28\%$, indicating the strong impact of degradation on charging behavior. SOC polynomials (22\%) capture nonlinear state-dependent dynamics, while CC-CV indicators (12\%) represent the phase transition at $80\%$ SOC. The individual feature analysis in Fig.~\ref{fig:feature_importance_b} further emphasizes the role of CV-phase modeling: tapering features (CV Taper², CV Taper³) collectively contribute 18\% of predictive power, confirming that modeling exponential power decay during CV charging is critical. SoH-related features (SoH², SOC×SoH, CV×SoH) also appear prominently among the top contributors, validating the need for degradation-aware feature construction. 

Overall, engineered nonlinear transformations, including polynomial terms, tapering indicators, and SoH interactions, account for 62\% of the total predictive power, explaining the 92.6\% accuracy improvement over linear models (28.4\% $\rightarrow$ 2.1\% MAPE). Temperature features contribute about $6\%$, indicating moderate sensitivity within the tested range (-10°C to 40°C).

\subsection{RL Training Stability}

\begin{figure}[t!]
    \centering
    \includegraphics[width=0.97\linewidth]{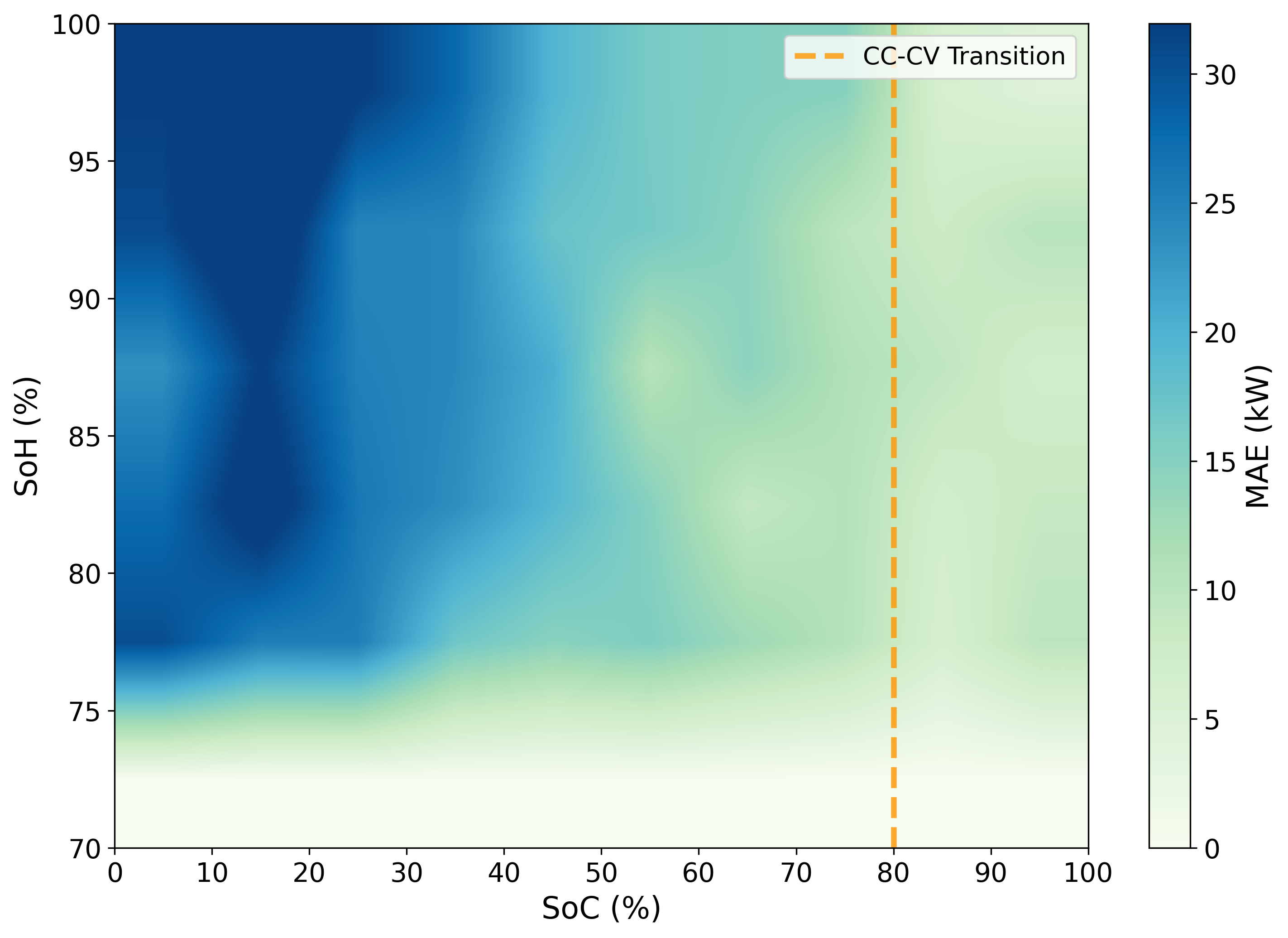}
    \caption{RL training stability over 10,000 episodes.}
    \label{fig:fig10}
\end{figure}

Fig.~\ref{fig:fig10} shows the RL training dynamics over 10,000 episodes. Episode rewards improve from -12 to -2.5 and stabilize after episode 6,000, while Q-values converge to the 15–18 range, indicating stable policy learning. The epsilon decay from 0.3 to 0.05 supports a smooth shift from exploration to exploitation. This heatmap shows the agent's prediction accuracy across charging scenarios, where the deep blue regions (high MAE) indicate initial difficulty at low SoC levels, where battery dynamics are most variable. The transition to lighter colors demonstrates learning convergence. In addition, the agent's low prediction errors across the CC-CV transition boundary validate the DQN architecture's capacity to capture nonlinear battery behavior, with near-uniform low MAE across the state space except in rare edge cases. This convergence demonstrates the RL model's ability to adaptively learn complex state-dependent dynamics through experience, achieving 1.6\% MAPE performance.

%%%%%%%%%%%%%%%%%%%%%%%%%%%%%%%%%%%%%%%%%%%%%%%%%%%%%%%%%%%%%%%%%%%%%%%%%%%%%%%%
\section{Conclusion}\label{Section 6}
In this paper, we presented a hybrid analytical--RL framework for EV charging time prediction that captures the nonlinear CC--CV dynamics while explicitly accounting for battery degradation. We formulated charging time as an SoH-dependent integral over the CC--CV trajectory and developed a physics-informed analytical model that provides accurate predictions even with limited training data. Building on this foundation, we introduced an RL framework based on a deep Q-network (DQN) that progressively improves prediction accuracy as additional charging-session data become available. We incorporated a progressive learning strategy to enable data-efficient training and robust adaptation to long-term battery aging. Extensive numerical studies demonstrated that the analytical model provides a strong baseline, achieving a $92.6\%$ improvement over linear methods, while the RL model offers an additional $23.8\%$ accuracy gain and $35\%$ better adaptation to SoH degradation. These results highlight the effectiveness of combining physics-based modeling with RL for accurate and resilient charging-time prediction. In future work, we will validate the proposed framework using commercial EV fleet data and extend the model to incorporate thermal dynamics and battery management systems.
\vspace{-2pt}

%%%%%%%%%%%%%%%%%%%%%%%%%%%%%%%%%%%%%%%%%%%%%%%%%%%%%%%%%%%%%%%%%%%%%%%%%%%%%%%%
\bibliography{ifacconf}  

\end{document}